\documentclass[12pt]{article}
 \usepackage{pstricks}
 \usepackage{epsfig}

\newcommand{\gr}{\tilde{G}}

\def\c{\chi}

\def\f{\phi}
\def\g{\gamma}

\def\j{\psi}

\def\m{\mu}
\def\n{\nu}

\def\p{\pi}

\def\r{\rho}
\def\s{\sigma}
\def\t{\tau}

\def\z{\zeta}

\def\G{\Gamma}

\def\L{\Lambda}
\def\O{\Omega}

\def\cm{{\cal M}}

\def\co{{\cal O}}

\def\bo{{\raise.15ex\hbox{\large$\Box$}}}               
\def\pr{\prod}                                          
\def\face{{\raise.2ex\hbox{$\displaystyle \bigodot$}\mskip-2.2mu \llap {$\ddot
        \smile$}}}                                      

\def\VEV#1{\left\langle #1\right\rangle}        
\def\leftrightarrowfill{$\mathsurround=0pt \mathord\leftarrow \mkern-6mu
        \cleaders\hbox{$\mkern-2mu \mathord- \mkern-2mu$}\hfill
        \mkern-6mu \mathord\rightarrow$}       
\def\dvec#1{\vbox{\ialign{##\crcr
        \leftrightarrowfill\crcr\noalign{\kern-1pt\nointerlineskip}
        $\hfil\displaystyle{#1}\hfil$\crcr}}}           

\def\beq{\begin{equation}}
\def\eeq{\end{equation}}

\def\beqx{\begin{displaymath}}
\def\eeqx{\end{displaymath}}

\def\beqa{\begin{eqnarray}}
\def\eeqa{\end{eqnarray}}
\def\NO{\nonumber}

\def\pl#1#2#3{Phys.~Lett.~{\bf B {#1}} (19{#2}) #3}
\def\np#1#2#3{Nucl.~Phys.~{\bf B {#1}} (19{#2}) #3}

\def\pr#1#2#3{Phys.~Rev.~{\bf D {#1}} (19{#2}) #3}

\def\ptp#1#2#3{Progr.~Theor.~Phys.~{\bf {#1}} (19{#2}) #3}

\def\sbs#1{{}_{\mbox{\scriptsize #1}}}          
\def\sbss#1{{}_{\mbox{\tiny #1}}}               
\newcommand{\gl}{\tilde{g}}                     
\newcommand{\wi}{\tilde{W_3}}                   
\newcommand{\bi}{\tilde{b}}                     
\newcommand{\hi}{\tilde{h_1^0}}                 
\newcommand{\hj}{\tilde{h_2^0}}                 
\newcommand{\qu}{\tilde{q}}                     

\catcode`@=11
\def\@citex[#1]#2{\if@filesw\immediate\write\@auxout{\string\citation{#2}}\fi
  \def\@citea{}\@cite{\@for\@citeb:=#2\do
    {\@citea\def\@citea{,\penalty\@m}\@ifundefined
      {b@\@citeb}{{\bf ?}\@warning
       {Citation `\@citeb' on page \thepage \space undefined}}%
\hbox{\csname b@\@citeb\endcsname}}}{#1}}

\def\citer{\@ifnextchar [{\@tempswatrue\@citexr}{\@tempswafalse\@citexr[]}}
\def\@citexr[#1]#2{\if@filesw\immediate\write\@auxout{\string\citation{#2}}\fi
  \def\@citea{}\@cite{\@for\@citeb:=#2\do
    {\@citea\def\@citea{--\penalty\@m}\@ifundefined
       {b@\@citeb}{{\bf ?}\@warning
       {Citation `\@citeb' on page \thepage \space undefined}}%
\hbox{\csname b@\@citeb\endcsname}}}{#1}}
\catcode`@=12

\begin{document}
\date{\mbox{ }}

\title{ 
{\normalsize     
DESY 98-066\hfill\mbox{}\\
September 1998\hfill\mbox{}\\}
\vspace{3cm}
\bf BARYON ASYMMETRY\\ AND DARK MATTER\\[8mm]}
\author{M.~Bolz, W.~Buchm\"uller, M.~Pl\"umacher\\
{\it Deutsches Elektronen-Synchrotron DESY, 22603 Hamburg, Germany}}
\maketitle

\thispagestyle{empty}

\vspace{2cm}
\begin{abstract}
\noindent
We study the implications of a large baryogenesis temperature,
$T_B = \co(10^{10}$~GeV), on the mass spectrum of superparticles in
supersymmetric extensions of the standard model. Models with a neutralino 
as lightest superparticle (LSP) are excluded. A consistent picture is 
obtained with the gravitino as LSP, followed by a higgsino-like neutralino 
(NSP). Gravitinos with masses from 10 to 100~GeV may be the dominant
component of dark matter.  
\end{abstract}

\newpage

\section{Introduction}

In the high-temperature phase of the standard model as well as its
supersymmetric extension, baryon number ($B$) and lepton number ($L$)
violating processes are in thermal equilibrium \cite{Kuz85}. 
As a consequence, asymmetries in baryon number and lepton number are
related at high temperatures,
  \beq
    \VEV{B}\sbs{T} = C \VEV{B-L}\sbs{T} = \frac{C}{C-1} \VEV{L}\sbs{T}\;,
    \label{eq:1}
  \eeq
where the constant $C=\co(1)$ depends on the particle content of the theory.
Hence, the cosmological baryon asymmetry can be generated from a
primordial lepton asymmetry produced by the out-of-equilibrium decay
of heavy Majorana neutrinos - a mechanism referred to as leptogenesis
\cite{Fuk86}. 

The natural theoretical framework for extensions of the standard model
with right-handed neutrinos are unified theories based on the gauge
group $SO(10)$. Corresponding models of leptogenesis \cite{Ghe93,Buc96}
can indeed explain the observed baryon asymmetry. 
Assuming a hierarchy of neutrino masses similar to the known mass
hierarchy of up-type quarks, one obtains a rather large temperature
for baryogenesis, $T_B\sim 10^{10}$~GeV \cite{Buc96}, which can be
reached in inflationary models after reheating \cite{Kol90}.
Particularly attractive are supersymmetric models of leptogenesis
\cite{Cam93,Plu97} for which a consistent picture including washout
processes and the generation of the initial equilibrium distribution
of the heavy Majorana neutrinos has been obtained \cite{Plu97}.

It is widely believed that the reheating temperature in the early universe 
cannot exceed $\co(10^{9}$~GeV) for a supersymmetric plasma \cite{Kol90} 
because of the `gravitino problem' \cite{Khl84,Ell84,Ell85}. 
In the high-temperature plasma a 
large number of gravitinos is generated. The late decay of unstable 
gravitinos after nucleosynthesis modifies the abundances of the light elements 
in a way which is incompatible with observation. Stable massive gravitinos, 
on the other hand, may overclose the universe. 
In the following we shall study this problem and investigate the
implications of a large baryogenesis temperature $T_B$ on the mass
spectrum of superparticles.

\section{Gravitino density}

The production of gravitinos ($\gr$) at high temperatures is dominated
by two-body processes involving gluinos ($\gl$). 
The corresponding cross sections have been considered for the two cases
$m_{\gr} > m_{\gl}$ \cite{Ell84,Mor93} and $m_{\gr} \ll m_{\gl}$ \cite{Mor93}.
In the Boltzmann equation, which is used to evaluate the gravitino density,
the thermally averaged zero-temperature cross sections enter. 

\begin{table}[h]
\begin{center}
\begin{tabular}{r|l|l}
 &\qquad process & \qquad$\bar{\sigma}_i$\\ \hline \hline
A & $g^a + g^b \rightarrow \gr + \gl^c$ & ${8\over 3}|f^{abc}|^2$\\ \hline 

B & $g^a + \gl^b \rightarrow \gr + g^c$ & $\left(8\ln{{s\over m^2}} -14\right)
                                          |f^{abc}|^2$\\ \hline

C & $g^a + \qu_i \rightarrow \gr + q_j$ & $4 |T^a_{ji}|^2$\\ \hline

D & $g^a + q_i \rightarrow \gr + \qu_j$ & $2 |T^a_{ji}|^2$\\ \hline

E & $q_i + \qu_j^* \rightarrow \gr + g^a$ & $4 |T^a_{ji}|^2$\\ \hline

F & $\gl^a + \gl^b \rightarrow \gr +\gl^c$ & $\left(16\ln{{s\over m^2}} -
                                     {92\over 3}\right)|f^{abc}|^2$\\ \hline

G & $\gl^a + q_i \rightarrow \gr + q_j$ & $\left(8\ln{{s\over m^2}}-16\right)
                                          |T^a_{ji}|^2$\\ \hline

H & $\gl^a + \qu_i \rightarrow \gr + \qu_j$ & $\left(8\ln{{s\over m^2}} 
                                            -14\right)|T^a_{ji}|^2$\\ \hline

I & $q_i + \bar{q}_j \rightarrow \gr + \gl^a$ & ${8\over 3} 
                                             |T^a_{ji}|^2$\\ \hline

J & $\qu_i + \qu_j^* \rightarrow \gr + \gl^a$ & ${16\over 3} 
                                             |T^a_{ji}|^2$

\end{tabular}
\medskip
\caption{\it Cross sections $\bar{\sigma}_i(s)$ for gravitino ($\gr$)
production in two-body processes involving left-handed quarks ($q_i$), 
scalar quarks
($\qu_i$), gluons ($g^a$) and gluinos ($\gl^a$). The cross sections are
given for the specified choice of colours and averaged over spins in the
initial state. $f^{abc}$ and $T^a_{ji}$ are the usual SU(3) colour matrices. }
\end{center}
\end{table}

For processes with a gluon in the t-channel a logarithmic collinear 
singularity appears which has been regularized \cite{Ell84,Mor93}
by introducing either a finite gluon mass or an angular cut around the 
forward direction. We have calculated all partial cross sections $\s_i$ 
using both methods. The logarithmically singular terms 
are universal whereas the finite parts depend on the cutoff procedure. 
For arbitrary gravitino masses and large centre-of-mass energies,
$s\gg m_{\gr}^2, m_{\gl}^2$, the different cross sections are given by 
\beq
\s_i(s) = {g^2\over 64\pi M^2}\left(1+{m_{\gl}^2\over 3 m_{\gr}^2}\right)
           \bar{\s}_i(s)\;,
\eeq
where $g$ is the QCD coupling,  
and $M=(8\p G\sbs{N})^{-\frac{1}{2}}\simeq2.4\cdot10^{18}$~GeV is the
Planck mass. For $m_{\gr} \ll m_{\gl}$ the production cross section is 
enhanced since the scattering amplitude for the Goldstino component of the
gravitino is inversely proportional to the supersymmetry breaking
scale, $\cm\propto \frac{1}{\L^2\sbss{SUSY}} \propto \frac{1}{m_{\gr}M}$.
Our results for the partial cross sections $\bar{\s}_i(s)$
are listed in the table. Here the collinear singularity has been regularized
by a finite gluon mass $m$, and the logarithmically singular and the constant
part are given for each process. The coefficients of the $\ln(s/m^2)$ terms
are in agreement with results obtained by Moroi \cite{Mor98}.

In a consistent finite-temperature calculation, which remains to be carried 
out, the logarithmic singularity has to be regularized by the relevant 
finite-temperature mass scale which is expected to be the plasmon mass, i.e.
$m \sim g(T) T$. We therefore estimate the gravitino production
rate by evaluating the thermal average of the universal $\ln(s/m^2)$ term
(processes B, F, G, H),
\beq
\s_{(L)} = {g^2\over 2\pi M^2}\eta(1+\eta)((N^2-1)C_A + 2n_f N C_F)
\left(1+{m_{\gl}^2 \over 3 m_{\gr}^2}\right) \ln{\left({s\over m^2}\right)}\;.
\eeq
Here we have summed over all colours and all spins in the initial state,
and included symmetry factors and a factor $\eta$ for each fermion in
the initial state. $C_A$ and $C_F$ are the usual colour factors for the
group SU(N) and $2n_f$ is the number of 
colour-triplet chiral multiplets, i.e. $2n_f=12$ in the MSSM. The 
corresponding thermally averaged cross section reads ($\eta=3/4$)
\beqa
C(T) &=& \VEV{\s_{(L)} v\sbs{rel}} \NO\\
&=& {21 g^2(T)\over 32\pi\z^2(3) M^2}((N^2-1)C_A + 2n_f N C_F)
\left(1+{m_{\gl}^2(T) \over 3 m_{\gr}^2}\right)\NO\\
&&\hspace{2cm}\left(\ln{1\over g^2(T)} + {5\over 2} + 2\ln{2} -2\g_E\right)\;,
\eeqa
where we have substituted $m$ by $g(T)T$.
We expect that the unknown constant part of the thermally averaged cross
section contributes to $C(T)$ about the same amount as the term proportional to
$\ln(1/g^2(T))$ (cf.~table 1).

The production cross section $C(T)$ enters in the Boltzmann equation, which 
describes the generation of a gravitino density $n_{\gr}$ in the thermal bath 
(cf.~\cite{Kol90}),
  \beq
    \frac{dn_{\gr}}{dt} + 3 H n_{\gr} = C(T) n\sbs{rad}^2\;.
    \label{eq:2}
  \eeq
Here $H(T)$ is the Hubble parameter and
$n\sbs{rad}=\frac{\z(3)}{\p^2}T^3$ is the number density of a relativistic 
bosonic degree of freedom. For QCD (N=3) one has
\beq\label{eq:3}
C(T) \simeq 10 {g^2(T)\over M^2}\left(1+{m_{\gl}^2(T) \over 3 m_{\gr}^2}\right)
\left(\ln{1\over g^2(T)} + 2.7\right)\;.
\eeq

From eqs.~(\ref{eq:2}) and (\ref{eq:3}) one obtains for the gravitino
density at temperatures $T<T_B$, assuming constant entropy, 
  \beq
    Y_{\gr}(T)\equiv\frac{n_{\gr}(T)}{n\sbs{rad}(T)}\simeq
    \frac{g_{\star S}(T)}{g_{\star S}(T_B)}
    \frac{C(T_B)n\sbs{rad}(T_B)}{H(T_B)}\;,
    \label{eq:5}
  \eeq
where $g_{\star S}(T)$ is the number of effectively massless degrees
of freedom \cite{Kol90}. For $T<1$~MeV, i.e. after nucleosynthesis, 
$g_{\star S}(T)=
2+\frac{21}{4}\left(\frac{T_{\n}}{T}\right)^3
=\frac{43}{11}$, 
and $g_{\star S}(T_B)=\frac{915}{4}$ in the MSSM.
For light gravitinos ($m_{\gr}\ll m_{\gl}(\m), \m\simeq 100$~GeV) one obtains
from eqs.~(\ref{eq:3})-(\ref{eq:5})
for the gravitino density and the contribution to $\O h^2$, 
  \beq
    Y_{\gr}\simeq 3.2\cdot 10^{-10}
    \left(\frac{T_B}{10^{10}\,\mbox{GeV}}\right)
    \left(\frac{100\,\mbox{GeV}}{m_{\gr}}\right)^2
    \left(\frac{m_{\gl}(\m)}{1\,\mbox{TeV}}\right)^2,
    \label{eq:6}
  \eeq

  \beqa
    \O_{\gr}h^2 & = & m_{\gr} Y_{\gr}(T) n\sbs{rad}(T) \r_c^{-1} \NO \\
    & \simeq & 0.60
    \left(\frac{T_B}{10^{10}\,\mbox{GeV}}\right)
    \left(\frac{100\,\mbox{GeV}}{m_{\gr}}\right)
    \left(\frac{m_{\gl}(\m)}{1\,\mbox{TeV}}\right)^2.    
    \label{eq:7}
  \eeqa
Here we have used $g(T_B)=0.85$; $\r_c=3H_0^2M^2$ is the critical energy
density, and $m_{\gl}(T)=\frac{g^2(T)}{g^2(\m)} m_{\gl}(\m)$. 
If one assumes that the running 
masses of gluino, wino and bino ($\bi$) unify at the GUT scale,
one has $m_{\gl}(\m)=\frac{3}{5}\frac{g^2(\m)}{\bar{g}^2(\m)} m_{\bi}(\m)$,
where $\bar{g}$ is the $U(1)_Y$-gauge coupling.

\section{Constraints from nucleosynthesis}

The primordial synthesis of light elements (BBN) yields stringent
constraints on the amount of energy which may be released after
nucleosynthesis by the decay of heavy nonrelativistic particles into
electromagnetically  and strongly interacting relativistic particles. These 
constraints have been studied in detail by several groups \cite{Ell92,Mor93}. 
Depending on the lifetime of the decaying particle $X$ its
energy density cannot exceed an upper bound. From fig.~3 in
\cite{Ell92} one reads off that one of the following conditions is
sufficient:
  \beqa
    \mbox{(I)} & m_X Y_X(T) < 4\cdot 10^{-10}\,\mbox{GeV}, &
    \t < 2\cdot 10^6\,\mbox{sec},
    \label{eq:8} \\
    \mbox{(II)} & m_X Y_X(T) < 4\cdot 10^{-12}\,\mbox{GeV}, &
    \t\,\,\mbox{arbitrary},
    \label{eq:9}
  \eeqa
where $Y_X(T) = n_X(T)/n\sbs{rad}(T)$.

Gravitinos interact only gravitationally. Hence, their existence leads
almost unavoidably to a density of heavy particles which decay after
nucleosynthesis. The partial width for the decay of an unstable
gravitino into a gauge boson $B$ and a gaugino $\bi$ is given by
\cite{Mor93} ($m_{\bi}\ll m_{\gr}$),
  \beq
    \G(\gr \rightarrow B\bi)  \simeq  
    \frac{1}{32\pi}\frac{m_{\gr}^3}{M^2} 
    \simeq  \left[ 4\cdot10^8\left( \frac{100\,\mbox{GeV}}{m_{\gr}}
    \right)^3\,\mbox{sec}\right]^{-1}.
    \label{eq:10}
  \eeq
If for a fermion $\j$ the decay into a final state with a scalar $\f$ in
the same chiral multiplet and a gravitino is kinematically allowed, 
the partial width reads, 
  \beq
    \G(\j \rightarrow \gr \f) =\G(\j \rightarrow \gr \f^*)
     \simeq \frac{1}{96\pi}\frac{m_{\j}^5}{m_{\gr}^2M^2}\;. 
    \label{eq:11}
  \eeq
Given these lifetimes and the mass spectrum of superparticles in the
MSSM one can examine whether one of the conditions (I) and (II)
on the energy density after nucleosynthesis is satisfied.

\section{Mass spectrum of superparticles}

Consider first a typical example of supersymmetry breaking masses in
the MSSM, $m_{\bi}<m_{\gr}\simeq 100\,\mbox{GeV}<m_{\gl}\simeq 500\,
\mbox{GeV}$, and $T_B \simeq 10^{10}$~GeV.
From eqs.~(\ref{eq:6}) and (\ref{eq:10}) we conclude $\t_{\gr}\simeq
4\cdot10^8$~sec, $m_{\gr}Y_{\gr}(T)\simeq 4\cdot10^{-9}$~GeV. 
According to condition (II) (\ref{eq:9}) this energy density
exceeds the allowed maximal energy density by 3 orders of
magnitude. This clearly illustrates the `gravitino problem'!

The existence of an unstable gravitino is inconsistent with a 
baryogenesis temperature $T_B$ as large as $10^{10}$~GeV. Consider first
condition (I) (\ref{eq:8}). To satisfy the lifetime constraint 
$\t<2\cdot10^6$~sec one needs, according to (\ref{eq:10}),
$m_{\gr}>600$~GeV. Eqs.~(\ref{eq:3}) and (\ref{eq:6}) then imply 
$m_{\gr}Y_{\gr}>2.3\cdot10^{-8}$~GeV, which exceeds the upper bound of 
condition (I) by 2 orders of magnitude. Condition (II) can also not be 
satisfied, since it would require an LSP mass below the experimental bounds. 
  \begin{figure}[h]
\begin{center}
    \vskip .1truein
    \centerline{\epsfysize=10cm {\epsffile{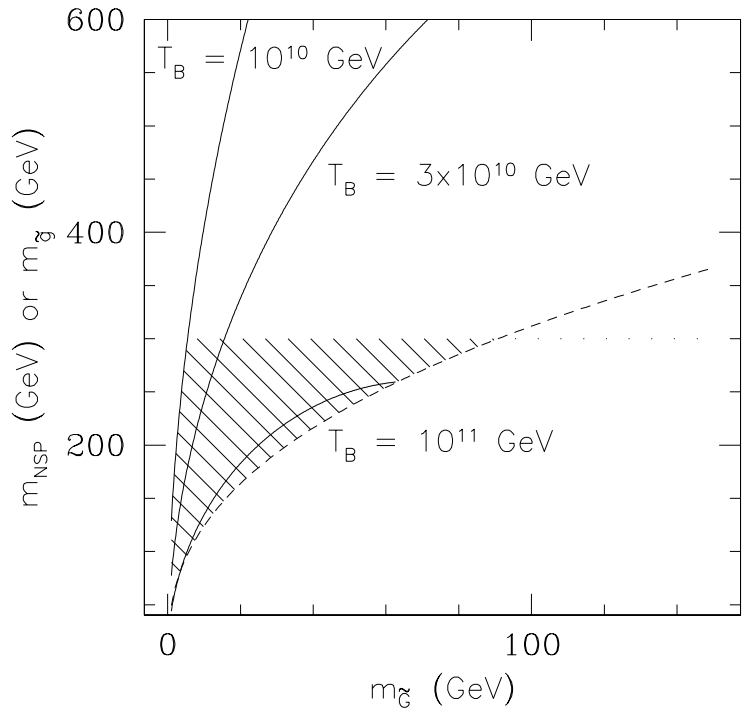}}}
    \vskip -0.3truein
\caption{\it Upper and lower bounds on the NSP mass as function of the
gravitino mass. The full lines represent the upper bound on the gluino 
mass $m_{\gl}(\m) > m_{NSP}$ for different reheating 
temperatures. The dashed line is the lower bound on $m_{NSP}$ which follows
from the NSP lifetime. A higgsino-like NSP with a mass in the shaded area
satisfies all cosmological constraints including those from primordial
nucleosynthesis.}
    \label{fig:1}
\end{center}  
\vspace{-1cm}
\end{figure}

Consider now the case in which the gravitino is the LSP, a possibility
previously discussed in \cite{Ell85,Ber91}. In this case
one has to worry about the decays of the next-to-lightest
superparticle (NSP) after nucleosynthesis. The lifetime constraint of
condition (I), $\t\sbs{NSP}<2\cdot10^6$~sec, yields a lower bound on
the NSP mass which depends on the gravitino mass $m_{\gr}$ 
(cf.~eq.~(\ref{eq:11}) and fig.~\ref{fig:1}).
For a large range  of parameters the NSP is a neutralino $\c$, 
i.e. a linear combination of higgsinos and gauginos,
  \beq
    \c = N_1 \bi + N_2 \wi + N_3 \hi + N_4 \hj\;. 
    \label{eq:12}
  \eeq
The NSP density after nucleosynthesis has been studied in great detail
by a number of authors \cite{Dre98}, since the density of stable neutralinos
would contribute to dark matter. 
The upper bound of condition (II) on the neutralino density,
$m_\c Y_\c (T)<4\cdot10^{-12}$~GeV, corresponds to the requirement 
$\O h^2<0.00008$. 

A systematic study of the neutralino density for a large range of the
MSSM parameter space has been carried out by Edsj\"o and Gondolo
\cite{Eds97}. For an interesting range of parameters, one finds a
value in the `cosmologically interesting' region $0.025<\O_\c h^2 <1$.
In general, however, $\O_\c h^2$ varies over eight orders of magnitude,
from $10^{-4}$ to $10^4$. For a large part of parameter space one finds 
$\O_\c h^2 < 0.025$. In particular this is the case for a higgsino-like 
neutralino, i.e. $Z_g=|N_1|^2+|N_2|^2<\frac{1}{2}$, in the mass range
$80\,\mbox{GeV}<m_\c <450\,\mbox{GeV}$ \cite{Eds97}. For these parameters 
neutralino pair annihilation into $W$ boson pairs is very efficient and one 
therefore obtains a small neutralino density. 

The bound $\O h^2<0.008$, which corresponds to the bound on the mass density 
$m_\c Y_\c (T)<4\cdot10^{-10}$~GeV of condition (I),
is satisfied for higgsino-like neutralinos
in the mass range $80\,\mbox{GeV}<m_\c <300\,\mbox{GeV}$ \cite{Gon98}.
We conclude that higgsino-like NSPs in this mass range and with a lifetime
$\t<2\cdot10^6$~sec are compatible with the constraints from
primordial nucleosynthesis.
Note that this is a sufficient, yet not necessary condition for satisfying
the bound $\O h^2<0.008$. Very small neutralino densities are also obtained
for other sets of MSSM parameters.  
  \begin{figure}[t]
\begin{center}
    \vskip .1truein
    \centerline{\epsfysize=10cm {\epsffile{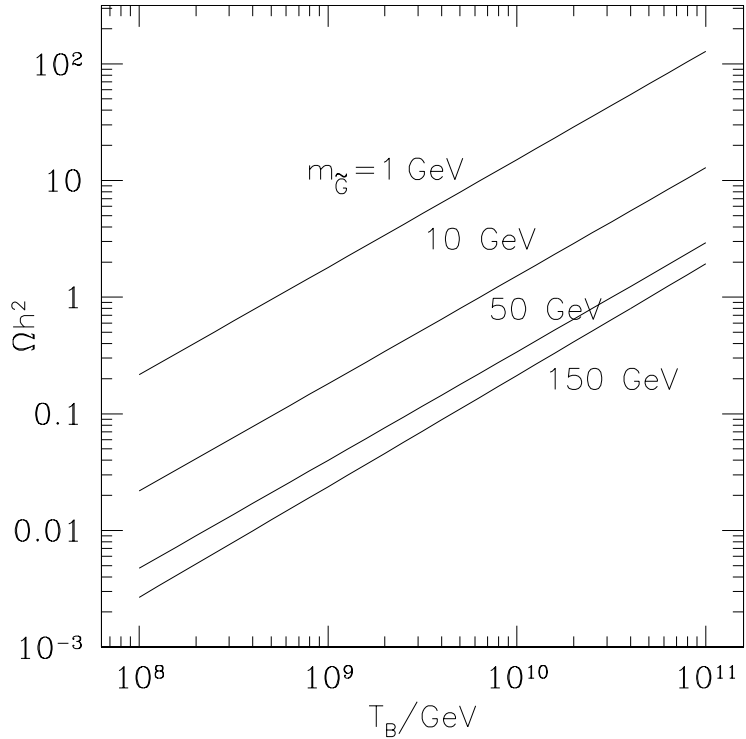}}}
    \vskip -0.1truein
    \caption{\it Contribution of gravitinos to the density parameter $\O h^2$
     for different gravitino masses $m_{\gr}$ as function of the reheating
     temperature $T_B$. The gluino mass has been set to $m_{\gl}(\m)=500$~GeV.}
    \label{fig:2}
\end{center}
\vspace{-1cm}
  \end{figure}

Finally, we have to discuss the necessary condition that gravitinos do
not overclose the universe,
  \beq
    \O_{\gr} h^2 < 1\;.
    \label{eq:13}
  \eeq
Since $m_{\gr}<m_\c <m_{\gl}$, the gravitino density is given by
eq.~(\ref{eq:7}). Hence, the condition (\ref{eq:13}) yields an upper bound on
the NSP mass $m_\c$ which depends on the gravitino mass and the baryogenesis 
temperature. The different constraints are summarized in fig.~\ref{fig:1} which
illustrates that for a wide range of MSSM parameters, where
  \beq
    m_{\gr}<m_\c <m_{\gl}\;,  \label{eq:14}
  \eeq
and $80\,\mbox{GeV}<m_\c <300\,\mbox{GeV}$, the baryogenesis
temperature may be as large as $\co(10^{10})$~GeV.\footnote{
The expression (\ref{eq:7}) for $\O_{\gr}
  h^2$ is not valid for very small gravitino masses suggested by
  gauge-mediated supersymmetry breaking and no-scale supergravity
  models. This possibility has been discussed in \cite{Bor96}.}

It is remarkable that for temperatures $T_B=10^8\dots 10^{11}$~GeV,
which are natural for leptogenesis, and for gravitino masses in the range
$m_{\gr}=10^0\dots 10^3$~GeV, which is expected for gravity induced
supersymmetry breaking, the relic density of gravitinos is cosmologically
important (cf.~fig.~\ref{fig:2}).
As an example, for $T_B\simeq10^{10}$~GeV, $m_{\gl}(\m)\simeq 500$~GeV, and
$m_{\gr}\simeq 50$~GeV, one has $\O_{\gr} h^2 \simeq 0.30$.\\

We would like to thank V.~Berezinsky, J.~Ellis, P.~Gondolo, A.~Jakov\'ac,
T.~Moroi, S.~Sarkar and F.~Vissani for valuable discussions and correspondence.


\newpage

\begin{thebibliography}{99}

\bibitem{Kuz85}
  V.~A.~Kuzmin, V.~A.~Rubakov, M.~E.~Shaposhnikov, \pl{155}{85}{36}

\bibitem{Fuk86}
  M.~Fukugita, T.~Yanagida, \pl{174}{86}{45}

\bibitem{Ghe93}
  T.~Ghergetta, G.~Jungman, \pr{48}{93}{1546}

\bibitem{Buc96} 
  W.~Buchm\"uller, M.~Pl\"umacher, \pl{389}{96}{73}

\bibitem{Kol90} 
  E.~W.~Kolb, M.~S.~Turner, {\it{The Early Universe}}
  (Addison-Wesley, Redwood City, CA, 1990)

\bibitem{Cam93}
  B.~A.~Campbell, S.~Davidson, K.~A.~Olive, \np{399}{93}{111};\\
  L.~Covi, E.~Roulet, F.~Vissani, \pl{384}{96}{169}

\bibitem{Plu97}
  M.~Pl\"umacher, \np{530}{98}{207}, {\tt hep-ph/9704231}

\bibitem{Khl84}
M.~Yu.~Khlopov, A.~D.~Linde, \pl{138}{84}{265} 

\bibitem{Ell84}
  J.~Ellis, J.~E.~Kim, D.~V.~Nanopoulos, \pl{145}{84}{181}

\bibitem{Ell85}
J.~Ellis, D.~V.~Nanopoulos, S.~Sarkar, \np{259}{85}{175}  

\bibitem{Mor93}
  T.~Moroi, H.~Murayama, M.~Yamaguchi, \pl{303}{93}{289};\\
  M.~Kawasaki, T.~Moroi, \ptp{93}{95}{879};\\
  T.~Moroi, Ph.D. thesis, {\tt hep-ph/9503210}

\bibitem{Mor98}
T.~Moroi, private communication

\bibitem{Ell92}
  J.~Ellis, G.~B.~Gelmini, J.~L.~Lopez, D.~V.~Nanopoulos, S.~Sarkar,
  \np{373}{92}{399}

\bibitem{Ber91}
V.~S.~Berezinsky, \pl{261}{91}{71}
  
\bibitem{Dre98}
  For a recent review and references, see
  M.~Drees, {\it Particle Dark Matter Physics: An Update}, APCTP 98-004, 
  {\tt hep-ph/9804231}

\bibitem{Eds97}
  J.~Edsj\"o, P.~Gondolo, \pr{56}{97}{1879}

\bibitem{Gon98}
P.~Gondolo, private communication

\bibitem{Bor96}
  S.~Borgani, A.~Masiero, M.~Yamaguchi, \pl{386}{96}{189}

\end{thebibliography}
\end{document}